\documentclass[aps,prb,twocolumn,superscriptaddress,floatfix]{revtex4}
\usepackage{graphicx}
\usepackage{psfrag}
\usepackage{color}
\usepackage{subfigure}
\usepackage{amsmath}
\definecolor{dred}{rgb}{0.7,0.0,0.0}
\allowdisplaybreaks 

\newcommand{\tlfese}{TlFeSe$_2$}

\begin{document}

\title{Magnetic states of the five-orbital Hubbard model for \\
one-dimensional iron-based superconductors}

\author{Qinlong Luo}

\affiliation{Department of Physics and Astronomy, University of
Tennessee, Knoxville, Tennessee 37996, USA} 
\affiliation{Materials Science
and Technology Division, Oak Ridge National Laboratory, Oak Ridge,
Tennessee 37831, USA}

\author{Kateryna Foyevtsova}
\author{German D. Samolyuk}
\author{Fernando Reboredo}
\affiliation{Materials Science
and Technology Division, Oak Ridge National Laboratory, Oak Ridge,
Tennessee 37831, USA}

\author{Elbio Dagotto}
\affiliation{Department of Physics and Astronomy, University of
Tennessee, Knoxville, Tennessee 37996, USA} 
\affiliation{Materials Science
and Technology Division, Oak Ridge National Laboratory, Oak Ridge,
Tennessee 37831, USA}

\date{\today}

\begin{abstract}
The magnetic phase diagrams of models for quasi one-dimensional 
compounds belonging to the iron-based superconductors family are presented.
The five-orbital Hubbard model and the real-space Hartree-Fock 
approximation are employed, supplemented by density functional theory to
obtain the hopping amplitudes. Phase diagrams are constructed varying
the Hubbard $U$ and Hund $J$ couplings and at zero temperature. 
The study is carried out at electronic
density (electrons per iron) $n = 5.0$, which is of relevance for the
already known  material TlFeSe$_2$, and also at $n = 6.0$,
where representative compounds still need to be synthesized. At $n = 5.0$ there 
is a clear dominance of staggered spin order along the chain direction.
At $n = 6.0$ and the realistic Hund coupling $J/U = 0.25$, the phase
diagram is far richer 
including a variety of ``block'' states involving ferromagnetic clusters
that are antiferromagnetically coupled, in qualitative agreement with
recent Density Matrix Renormalization Group calculations for the three-orbital
Hubbard model in a different context. These block states arise from 
the competition between ferromagnetic order (induced by double exchange, 
and prevailing at large $J/U$) 
and antiferromagnetic order (dominating at small $J/U$).
The density of states and orbital compositions of the many phases are also
provided.
\end{abstract}

\maketitle

\section{Introduction}

The theoretical study of high critical temperature 
iron-based superconductors\cite{johnston2010} 
was initially centered on the concept of Fermi surface (FS) nesting, within a 
framework that assumes a weak
on-site Hubbard $U$ coupling strength. However, a variety of recent
experiments have indicated that there are serious 
deviations from this simplistic scenario.\cite{dai2012}
For example, robust local moments have been found at room temperature\cite{local}
that are incompatible with a weak coupling perspective where both
the local moments and long-range magnetic order should develop simultaneously upon
cooling.  The existence of nematic states\cite{davis} and orbital-independent
superconducting gaps\cite{shimojima} are other experimental observations that 
cannot be rationalized merely by FS nesting. 
At present, from the results of several experiments 
and calculations there is a convergence to the notion that 
these materials are in an ``intermediate'' coupling
regime with regards to the strength 
of the Hubbard $U$ 
interaction.\cite{dai2012,basov,rong2009,luo-neutrons,DFT+DMFT,Kotliar}
This coupling regime certainly represents a challenge to theorists
because standard many-body techniques are not sufficiently 
developed to study with accuracy 
the difficult intermediate $U$ regime. 
In addition, the nature of the problem requires using
complex multiorbital Hubbard models that are considerably more difficult 
to study than the simpler one-orbital models used for cuprates.

Adding to the complexity that emerges from these recent investigations
is the discovery of insulating states
in alkali metal iron selenides, with a chemical composition very close
to those of superconducting samples.\cite{dagotto2013}
In particular, a novel magnetic state of the iron superconductors
was reported for the intercalated iron selenide K$_{0.8}$Fe$_{1.6}$Se$_2$, with
iron vacancies in a $\sqrt{5}\times\sqrt{5}$ arrangement.
Neutron scattering studies\cite{bao} of this (insulating) compound
revealed an unusual spin arrangement involving
2$\times$2 iron blocks with their four spins
ferromagnetically ordered, supplemented by an in-plane 
antiferromagnetic coupling of these
2$\times$2 magnetic blocks. K$_{0.8}$Fe$_{1.6}$Se$_2$
also has a large ordering temperature
and large individual magnetic moments $\sim$3.3~$\mu_B$/Fe.  Note that phase
separation tendencies have also been reported in this type
of materials,\cite{ricci} 
and for this reason 
the relation between the exotic magnetic order involving ferromagnetic blocks 
described above and
superconductivity is also a matter of intense discussion.

These developments suggest that new insights
into iron pnictides and chalcogenides could
be developed if the iron spins are arranged differently than in the nearly square
geometry of the  FeSe layers. For this reason considerable interest was generated
by recent studies\cite{caron1,petrovic,caron2,nambu,Sefat}
of BaFe$_2$Se$_3$ (123) that contains substructures
with the geometry of two-leg ladders,
similarly as those widely discussed before in the context of Cu-based
superconductors.\cite{ladders-original,dagotto-rice,dagotto-ladder}
Neutron diffraction studies of the 123-ladder at low temperature~\cite{caron1,nambu}
reported a magnetic order involving blocks
of four iron atoms with their moments aligned, coupled antiferromagnetically
along the ``leg'' ladder direction.
This ``Block-AFM'' state is similar
to that discussed before for K$_{0.8}$Fe$_{1.6}$Se$_2$, with the iron vacancies.
When the 123-ladder material is hole doped with K,
the magnetic state evolves from the Block-AFM state
to a spin state (called CX) for the case of KFe$_2$Se$_3$
with ferromagnetically align spins along the rungs,
coupled antiferromagnetically along the legs of the ladders.\cite{caron2}
Adding to the importance of these novel ladder materials, a single layer of
alkali-doped FeSe with the geometry of weakly coupled
two-leg ladders was recently reported to be superconducting.\cite{SCladder}
These Fe-based ladders provide
a simple playground where superconductivity and magnetism
can be explored, in a geometry that simplifies the theoretical
studies due to its quasi one-dimensionality. 

In recent real-space Hartree-Fock (HF) calculations of ladder models,\cite{luo2013-ladders}
the 2$\times$2 Block-AFM state was indeed found to be stable in 
a robust region of parameter space, after
a random initial magnetic configuration 
was used as the start of the iterative process
to avoid biasing the results.
The other recently observed\cite{caron2} CX-state was also found in
the theoretical phase diagrams. Several other
competing states, that could be stabilized in related compounds
via chemical doping or high pressure, were also identified.\cite{luo2013-ladders}
The variety of possible magnetic states is remarkable,
revealing  an unexpected level of complexity in these systems. A technical aspect
important for our purposes is that the study in Ref.~\onlinecite{luo2013-ladders} showed
that the use of the HF technique, which is a crude approximation particularly in
a low-dimensional geometry, did capture
the essence of the magnetic order in ladders at least when compared with Density
Matrix Renormalization Group calculations. This previous agreement allow us to
be confident that a HF study in one dimension, as presented below, may still capture the
dominant correlations at short distances in the phase diagrams.

\begin{figure}[tb]
\begin{center}
\includegraphics[trim = 0mm 0mm 0mm 0mm, clip,width=1.0\columnwidth]{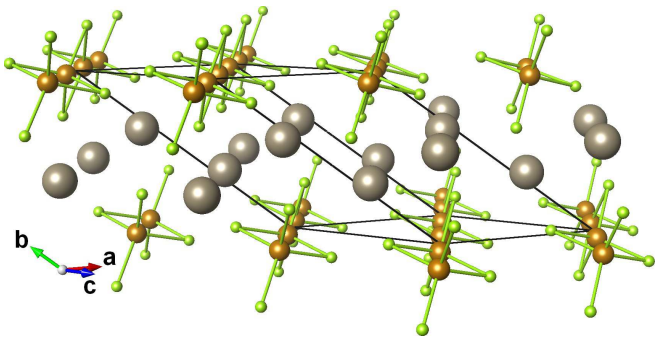}
\end{center}
\caption{Crystal structure of the one-dimensional compound {\tlfese}. The spheres
with small, medium, and large size represent Se, Fe, and Tl atoms,
respectively. The primitive unit cell and its lattice vectors are also shown.}
\label{figure.structure}
\end{figure}

In addition to the BaFe$_2$Se$_3$ compound already mentioned,
there is another group of Fe-selenide materials that also
display quasi one-dimensional characteristics but in this case simply
involving chains as opposed to ladders.
A typical representative is  TlFeSe$_2$. Its crystal structure is
in Fig.~\ref{figure.structure}, and it contains
dominant substructures in the form of weakly coupled chains.\cite{veliyev,seidov}
Iron is in a state Fe$^{3+}$, which
corresponds to $n=5$ for the electronic population
of the $3d$ Fe orbitals ($n$ is the average 
number of electrons in the $3d$ shell of the iron atoms).
TlFeSe$_2$ is believed to be antiferromagnetic, based on magnetic
susceptibility and transport measurements, with a characteristic
temperature $\sim 290$~K where short-range order
along the chains develops,\cite{veliyev} 
followed upon cooling by the stabilization of three-dimensional
long-range order at a lower temperature $\sim 14$~K.
As a consequence, over a wide temperature range the material behaves as nearly-independent
quasi-one dimensional chains. 
There are other compounds with similar structures,\cite{veliyev,seidov,K-1D} 
such as TlFeS$_2$ and KFeSe$_2$. In particular, TlFeS$_2$ has been studied
with neutron diffraction and 
it is also believed to be antiferromagnetic at low temperatures.\cite{weiz}
In spite of all this progress, 
the microscopic details of the spin order arrangements are unknown.

It is interesting to recall that in the
context of the Cu-based superconductors the study of materials that contain dominant 
one-dimensional 
chain substructures, such as SrCuO$_2$~\cite{srcuo2} and Sr$_2$CuO$_3$,\cite{sr2cuo3}
as well as the already mentioned
two-leg ladders,\cite{ladders-original,dagotto-rice,dagotto-ladder}
led to considerable advances in the understanding of cuprates. 
For all these reasons, in the present effort the theoretical study 
of the magnetic states of
models for one-dimensional iron-based superconductors will be initiated. Our aim
is to present a HF comprehensive study of multiorbital Hubbard models
in one dimension to predict qualitatively the dominant magnetic states in this context,
thus guiding future neutron scattering experiments. The previously discussed success 
in the use of the HF approximation when applied to ladders give us confidence that
at least qualitatively 
this crude approximation may still reveal the dominant magnetic tendencies
in one-dimensional systems. Of course, it is important to realize that
long-range order in the HF sense may in practice only correspond to 
dominant power-law behavior in
a particular ordering channel for the one-dimensional systems of interest here, 
or merely short-range order tendencies.

The electronic density $n=5$ of relevance for TlFeSe$_2$ will be studied here,
but other electronic densities, such as $n=6$ which is of relevance in
two-dimensional pnictides, will also be analyzed. This extension in the
range of $n$ is in anticipation of  the possible synthesis of novel 
one-dimensional materials with other values of $n$ in the future. 
In general terms, our overarching goal
is to motivate further experimental and theoretical efforts in the study 
of one-dimensional iron-based materials since
their theoretical analysis is simpler 
than in two dimensions and considerable insight 
on these exotic compounds can be gained by their detailed study and
subsequent comparison with experiments.

The organization of the manuscript and our main results are the following.
In Sec.~II, details of the density functional theory calculation are 
described, together with the five-orbital Hubbard 
model and the many-body technique employed. Section~III contains the phase
diagrams at both densities $n=5$ and $6$, in one and in anisotropic two dimensions.
Section~IV contains the density of states and orbital composition. Finally,
the conclusions are provided in Sec.~V. The details of the hopping amplitudes
are provided in the Appendix.

\section{Model and Technique}

The theoretical work presented in this section consists of two parts. First,
using $ab$-$initio$ techniques the band structure of TlFeSe$_2$ 
was calculated, and tight-binding hopping amplitudes were deduced.
Second, using the band structure information, Hubbard models involving
the five $3d$ orbitals of Fe were constructed and studied with
HF approximations. The detail is the following:

\subsection{Band structure calculation}


The electron Fe-Fe hopping amplitudes corresponding to the five Fe~$3d$ orbitals 
in TlFeSe$_2$ were calculated using two independent density functional theory (DFT) approximations.  
In the first one, the all-electron linearized augmented plane wave 
method, as implemented in WIEN2k~\cite{w2k} with the Perdew-Burke-Ernzerhof approximation
for the exchange-correlation functional,\cite{PBE} was employed to obtain the electronic structure 
of TlFeSe$_2$, which was subsequently projected onto a Wannier functions basis.\cite{Aichhorn09} 
The second calculation has been carried out in the frame of the local spin density approximation
with the Perdew and Zunger (PZ) exchange-correlation functional,\cite{PZ} using the 
plane wave pseudopotential method as implemented in the PWSCF code of the Quantum 
ESPRESSO (QE) distribution.\cite{QE} The ultrasoft pseudo-potential\cite{USPSPOT}  
optimized in the RRKJ scheme\cite{RRKJ} was used (Se.pz-n-rrkjus.UPF, Fe.pz-spn-rrkjus.UPF from 
the QE pseudopotentials database were employed, while the Pb.pz-dn-rrkjus.UPF pseudopotential was used as 
a prototype for the Tl pseudopotential). Furthermore, the result was used to calculate 
Maximally Localized Wannier functions~\cite{MLWF} as it is implemented in the Wannier90 
distribution.\cite{W90} Since the difference between hopping amplitudes obtained 
by these two methods does not exceed a few percent, for simplicity 
below only the WIEN2k results will be used.

The DFT calculations presented here 
are based on the crystal structure refined by Klepp and Boller.\cite{Klepp79}
{\tlfese} belongs to the $C2/m$ space group with a monoclinic centered unit cell
(Fig.~\ref{figure.structure}). 
The edge-sharing FeSe$_4$-tetrahedra form layers of parallel chains
with slightly alternating nearest-neighbor Fe-Fe bond distances
in the $ac$-plane of the primitive unit cell, separated by Tl atoms; 
each primitive unit cell contains two formula units.

%

According to the DFT calculations, the Fe and Se states are strongly hybridized,
with the Fe~$3d$ character 
prevailing between -1.8 and 1.5~eV (Fermi level is at 0~eV). 
This energy range was used to construct the
projected Fe~$3d$ Wannier functions and calculate the corresponding hopping integrals.
The final tight-binding (TB) model, used in our model 
calculations, includes all hoppings between the Fe pairs in the
plane of Fe chains within the separation range of 8.24~\AA. 
After the model is constructed, different electronic densities will simply
be reached by changing a chemical potential.

The DFT-calculated values of some of the 
hopping amplitudes are in the Appendix. 
As expected, the nearest-neighbor hoppings along the Fe-chains are dominant, 
with the largest absolute value being approximately 0.43~eV. 
However, in order to reproduce accurately the fine details of the band structure 
it is necessary to consider quite a number of
longer-ranged hoppings in other directions as well. Thus, a judicious truncation
of the hopping range (details presented below) 
 must be carried out to render practical the HF approximation 
described below.


\subsection{Hubbard model setup and many-body technique}

A multi-orbital Hubbard model will be employed in this study, based exclusively on 
the Fe $3d$ electrons. Our emphasis will be on the magnetic states obtained 
by varying the coupling parameters. 
This Hubbard model includes all five 
Fe $3d$ orbitals $\{d_{z^2},d_{x^2-y^2},d_{xy},d_{xz},d_{yz}\}$, 
which are widely believed to be the most 
relevant orbitals 
at the Fermi surface for the iron-based superconductors, in agreement also
with the band structure calculations of the previous subsection.

In real space, this multi-orbital Hubbard model includes a tight-binding term defined as:
\begin{eqnarray}\label{E.H0k}
H_{\rm TB} = \sum_{<\mathbf{i,j}>} \sum_{\alpha,\beta,\sigma}
t^{\alpha\beta}_{ij} (c^\dagger_{\mathbf{i},\alpha,\sigma}
c^{}_{\mathbf{j},\beta,\sigma} + h.c.),
\end{eqnarray}
where $c^\dagger_{\mathbf{i},\alpha,\sigma}$ creates an electron with spin
$\sigma$ in the orbital $\alpha$ at site $\mathbf{i}$, and
$t^{\alpha\beta}_{ij}$ refers to the tunneling amplitude of electron hoppings
from orbital $\alpha$ at site $\mathbf{i}$ to orbital $\beta$ at site
$\mathbf{j}$. The Coulombic interacting portion of the multi-orbital Hubbard Hamiltonian is given by:
\begin{equation}\begin{split}  \label{eq:Hcoul}
  H_{\rm int}& =
  U\sum_{{\bf i},\alpha}n_{{\bf i},\alpha,\uparrow}n_{{\bf i},
    \alpha,\downarrow}
  +(U'-J/2)\sum_{{\bf i},
    \alpha < \beta}n_{{\bf i},\alpha}n_{{\bf i},\beta}\\
  &\quad -2J\sum_{{\bf i},\alpha < \beta}{\bf S}_{\bf{i},\alpha}\cdot{\bf S}_{\bf{i},\beta}\\
  &\quad +J\sum_{{\bf i},\alpha < \beta}(d^{\dagger}_{{\bf i},\alpha,\uparrow}
  d^{\dagger}_{{\bf i},\alpha,\downarrow}d^{\phantom{\dagger}}_{{\bf i},\beta,\downarrow}
  d^{\phantom{\dagger}}_{{\bf i},\beta,\uparrow}+h.c.),
\end{split}\end{equation}
where $\alpha,\beta=1, 2, 3, 4, 5$ denote the Fe $3d$ orbitals, ${\bf S}_{{\bf i},\alpha}$
($n_{{\bf i},\alpha}$) is the spin (electronic density) of orbital $\alpha$ at site
${\bf i}$ (this index labels sites of the square lattice defined by the irons), 
and the standard relation $U'=U-2J$ between the Kanamori parameters
has been used.
The first two terms give
the energy cost of having two electrons located in  the same orbital or in
different orbitals, both at the same site, respectively. The second line contains the Hund's
rule coupling that favors the ferromagnetic (FM) alignment of the spins in
different orbitals at the same lattice site. 
The ``pair-hopping'' term is in the third line and its coupling is equal to $J$ by symmetry.

In this effort, the Hartree-Fock approximation will be applied to the Coulombic  
interaction, restricted to act only ``on site'', to investigate 
the ground state properties. This HF approximation results in a HF Hamiltonian 
containing several unknown HF expectation values (due to the many combinations of orbitals
that can be made). All these expectation values need to be 
optimized by a self-consistent iterative numerical process. 
The expectation values are assumed independent from site to 
site in the numerical procedure, allowing 
the system to select spontaneously the state that minimizes the HF energy, thus
reducing the bias into the calculations. To start the self-consistent iteration 
process, first all the HF expectation values  are set to 
some randomly chosen numbers, defining a random initial state. 
These random-start process 
tends to be time consuming and at the end of the iterations
states are obtained that typically display close, but not perfect, regularity.
By hand, these imperfections are removed and then those regular states 
are used again as initial states in the HF iterative process
where they are further optimized.

At the end of this tedious procedure, the ground states are selected 
by comparing the final ground energies after convergence. As discussed before, 
the tight-binding hopping amplitudes are obtained from the DFT calculation. However, 
for comparison purposes results using the hoppings from Ref.~\onlinecite{Graser08},
which were deduced in a quite different context, 
will be used as well to judge how robust our HF results phase diagrams are.
Our conclusions are that the essence of the phase diagrams do not depend much
on the details of the hoppings.

The HF calculations reported here are carried out using three different lattices: 
(1) Our most important results are obtained on $32\times1$ clusters, 
namely on one-dimensional single chains. Here, only the hopping 
amplitudes along the Fe-chain direction (See Fig.~1 in the Appendix)
are employed, including 
$t^{11}_{[0,0]}$, $t^{12}_{[0,0]}$, $t^{12}_{[-1,1]}$, $t^{11}_{[-1,1]}$, 
$t^{11}_{[1,-1]}$, $t^{12}_{[1,-1]}$ and $t^{12}_{[-2,2]}$. 
The ratio of hopping amplitudes at maximum and minimum distances is 
$t^{12}_{[0,0]} / t^{12}_{[-2,2]} \approx 12$. 
(2) Also $16\times1$ one-dimensional single chains were used but
employing the hopping amplitudes from Ref.~\onlinecite{Graser08}. 
The ratio 
of hopping amplitudes at maximum and minimum distances 
is about $12$ in this case as well. 
(3) To develop at least a qualitative idea of the magnetic coupling between
the one-dimensional dominant structures, $8\times8$ two-dimensional square lattices
were also studied. 
All the hopping amplitudes presented in the Appendix were employed in this case.

Periodic boundary conditions (PBC) are used here for all the calculations.
In the self-consistent iterative process, the criteria of convergence 
is set so that the changes of the HF expectation values are less 
than $10^{-4}$. Under this criteria, 
the typical number of iterations can be as large as $10,000$ for the
two-dimensional calculations and as large as $5,000$ 
for the one-dimensional calculations,
particularly for the case of random starting configurations.

\section{Results}

\subsection{Electronic density $n$=5 with (truncated to 1D) TlFeSe$_2$ hoppings}

\begin{figure}[thbp]
\begin{center}
\includegraphics[width=9.6cm,clip,angle=0]{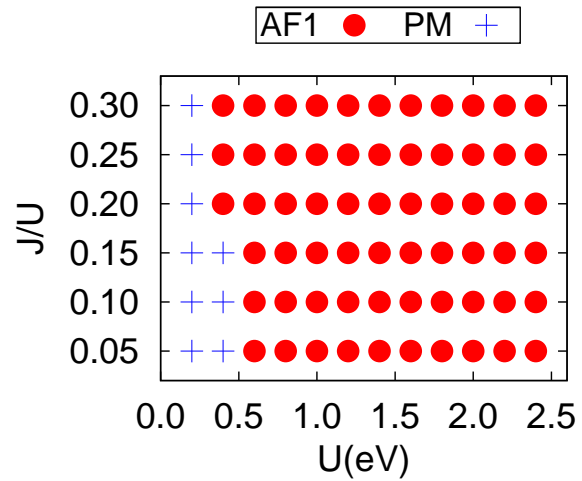}
\caption{Phase diagram obtained using a $32 \times 1$ 
one-dimensional lattice, employing the real-space
HF technique at a fixed electronic density $n = 5.0$. 
AF1 simply denotes a staggered AFM state, namely a pattern of spins 
$\uparrow \downarrow \uparrow \downarrow$ [see Fig.~\ref{PD1d}(b)].
The ``Paramagnetic'' (PM) state was 
defined with the following convention: 
the order parameter $m$ is smaller than 
a cutoff value chosen as $5\%$ of the saturated value 
for the same state at larger $U$ couplings. 
The hopping amplitudes used for this chain are in
the hopping matrices along the Fe-chain 
direction shown in the Appendix section, 
including $t^{11}_{[0,0]}$, $t^{12}_{[0,0]}$, $t^{12}_{[-1,1]}$, $t^{11}_{[-1,1]}$, 
$t^{11}_{[1,-1]}$, $t^{12}_{[1,-1]}$ and $t^{12}_{[-2,2]}$.
The bandwidth at $U=J=0$ is $W = 2.36$~eV.
}
\label{PDn51d}
\end{center}
\end{figure}

Let us start our description of the main results with the case of electronic
density $n=5$, corresponding to TlFeSe$_2$, and using first a one-dimensional
system containing 32 sites, and employing the real-space HF technique described
in the previous section. Increasing the Hubbard coupling $U$ to move beyond
the paramagnetic regime and for all the
Hund couplings $J$ analyzed here, there is only one magnetic state in the entire
phase diagram, namely the staggered $\uparrow \downarrow \uparrow \downarrow$ state
denoted as AF1 in Fig.~\ref{PDn51d}. This type of magnetic order is quite natural
in an electronic density regime with one electron per orbital at every site.

It should be remarked that there are two-dimensional analogs of the case described
here that also have the same staggered magnetic order. 
To be more precise, in recent efforts
a new avenue of research in iron-based superconductors has been expanding.
It consists of replacing entirely Fe by Mn or other $3d$ transition elements. 
The average electronic population of the $3d$ orbitals of Mn is $n = 5$,
different from the $n=6$ of Fe, but the crystal structures are similar as in
layered iron pnictides. As example, in the case of the 100\% replacement of Fe by Mn,
the compound BaMn$_2$As$_2$ was found to develop a G-type 
AFM state with staggered
spin order, a N\'eel temperature of 625~K, and a magnetic moment
of 3.88$\mu_B$/Mn at low temperatures.\cite{BMA} The G-type AFM order 
appears to be very robust, as recent investigations of 
Ba$_{1-x}$K$_x$Mn$_2$As$_2$ have shown.\cite{persistence} Again, this state
emerges naturally from the population $n=5$ at each Mn atom, namely
one electron per $3d$ orbital. For the same reason, it is natural that in our
one-dimensional case at $n=5$ a similar dominance of staggered spin order is found.

\subsection{Magnetic phase diagram for chain systems 
at electronic density $n$=6}

\begin{figure}[thbp]
\begin{center}
\includegraphics[width=8.0cm,clip,angle=0]{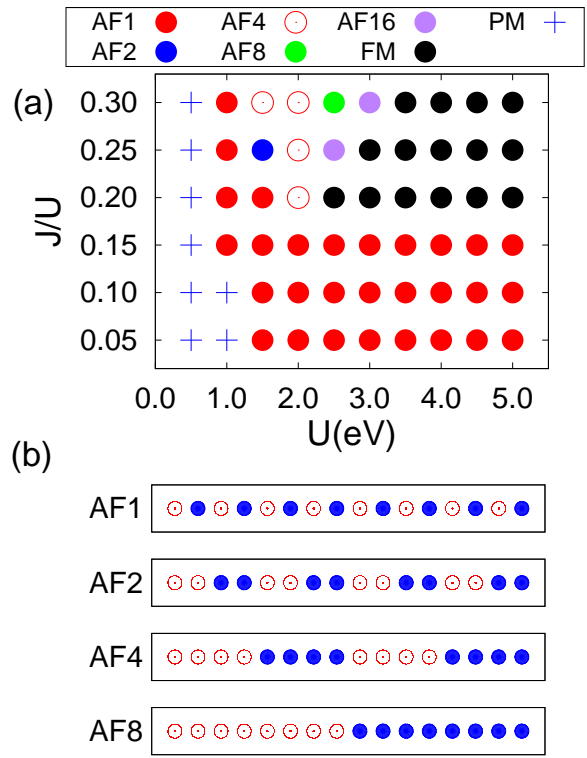}
\caption{(a) Phase diagram obtained 
using a  $32 \times 1$ one-dimensional lattice, employing real-space
HF techniques at a fixed electronic density $n = 6.0$. 
The ``Paramagnetic'' (PM) state was 
defined with the following convention: 
the order parameter $m$ is smaller than a cutoff value chosen as $5\%$ of the saturated value 
for the same state at larger $U$ couplings. 
The hopping amplitudes used for this chain are the hopping matrices along the Fe-chain 
direction (see Appendix section), 
including $t^{11}_{[0,0]}$, $t^{12}_{[0,0]}$, $t^{12}_{[-1,1]}$, $t^{11}_{[-1,1]}$, 
$t^{11}_{[1,-1]}$, $t^{12}_{[1,-1]}$ and $t^{12}_{[-2,2]}$.
The bandwidth at $U=J=0$ is $W = 2.36$~eV.
(b) Sketches of the states found in the phase diagram shown in (a), 
and their labeling convention.
}
\label{PD1d}
\end{center}
\end{figure}

The phase diagrams become far richer when the electronic density is changed
away from $n=5$.
Figure~\ref{PD1d} contains results that are expected to apply to materials with
electronic density $n=6.0$ that still
have not been synthesized to our knowledge. 
As in the previously described results, in this subsection the case of an
isolated chain is investigated since the hoppings perpendicular
to the chains are first neglected. 
In TlFeSe$_2$ there is a robust range of temperatures where
the physics is expected to be dominated by the one-dimensional structures, 
and it is in this regime that the present results are the most important 
if such a broad temperature window for
one dimensionality also exists in the yet-to-be-made $n = 6$ materials. 
In the absence of concrete information about possible $n=6$ chain compounds,
here the same hoppings as for the $n=5$ case of TlFeSe$_2$ case are used.

Figure~\ref{PD1d}~(a) contains the phase diagram 
varying the Hubbard $U$ and Hund $J$ couplings
at electronic density $n=6$. The complexity of the phase diagram, 
far richer than for the case $n=5$, 
is clear. The detail of the magnetic patterns in the many competing states 
is in Fig.~\ref{PD1d}~(b). At small $U$, Fig.~\ref{PD1d}~(a) has a 
paramagnetic state as expected. Increasing $U$, the spin staggered 
AF1 state is stabilized. This
occurs even at large values of $U$ as long as $J/U$ remains at 0.15 or below. 
However, in the more realistic regime of $J/U \sim 0.25$ a variety 
of competing states emerge. In this regime,
with further increasing $U$ eventually a ferromagnetic state dominates. This
large $J$ and $U$ regime resembles the physics of doped manganites, where the
double exchange mechanism generates the alignment of the spins.
In the intermediate regions
of $U/W \sim 0.5-1.0$, a cascade of transitions is present 
involving intermediate states with ferromagnetic
blocks, that are antiferromagnetically coupled. This is in good qualitative
agreement with recent investigations by Rinc\'on {\it et al.} 
using the Density Matrix Renormalization
Group (DMRG) technique applied to a 
three-orbital Hubbard model unrelated to TlFeSe$_2$.\cite{rincon}
 
If candidate one-dimensional selenides with $n=6.0$ 
are synthesized in the future, 
neutron scattering experiments will easily
pick up the dominant periodicity based on 
the position of the magnetic peaks in diffraction experiments.
However, it is very difficult to predict, even using $ab$-$initio$ 
calculations, where potentially stable materials
with $n=6.0$ will be precisely located in the phase diagram. 
But the concrete prediction of this model Hamiltonian calculations is
that the magnetic state must be one of those contained in the 
phase diagram of Fig.~\ref{PD1d} at intermediate $U/W$
and for $J/U \sim 0.25$. The complexity of these magnetic 
states provides motivation for the
search of materials with quasi one-dimensional structures 
and electronic density $n \sim 6.0$.

\subsection{Electronic density $n$=6 with (truncated to 1D) hoppings
from layered pnictides}

\begin{figure}[thbp]
\begin{center}
\includegraphics[width=9.0cm,clip,angle=0]{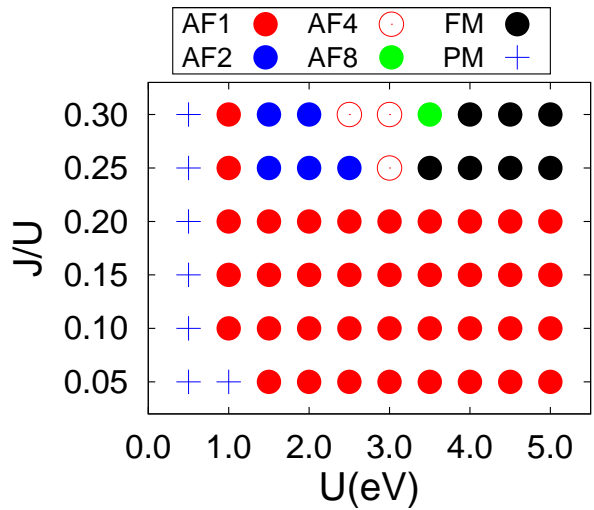}
\caption{Phase diagram for a $16 \times 1$ one-dimensional lattice, using the
HF approximation at fixed $n = 6.0$. The hopping amplitudes are taken from 
Ref.~\onlinecite{Graser08}, by truncating those hoppings to one dimension. 
The definition of ``Paramagnetic'' (PM) state is 
the same as in Fig.~\ref{PD1d}. The bandwidth at $U=J=0$ is $W = 2.2$~eV.
The notation for the many states was explained in previous figures.}
\label{PDGraser}
\end{center}
\end{figure}

With the only purpose of finding out how robust the previous results are,
a real-space HF phase diagram was constructed at $n=6$ for the case of hoppings 
that were actually calculated
in a different context, i.e. layered pnictides. Those hoppings were simply truncated
to a one-dimensional chain along one of the directions of the layers. Again,
there is no direct physical motivation for this portion of the 
calculation but merely the curiosity to gauge
how different the results would be by choosing quite a different set of hopping. 
To our surprise, the phase diagram at $n=6$ 
shown in Fig.~\ref{PDGraser} is very similar to that
in Fig.~\ref{PD1d}. 
Similarly as in Fig.~\ref{PD1d},
a paramagnetic phase exists at small $U$, a spin staggered state AF1 dominates 
at intermediate and large $U$
and $J/U$ not larger than 0.20, and a ferromagnetic state is stable 
at realistic $J/U$ and large $U$. In between, in the regime of $U/W$ between
0.5 to 1.5, approximately, a transition from AF1 to FM via intermediate Block
states with an increasing size of the FM blocks is obtained. Thus, in
spite of the use of quite different hoppings, the essential aspects 
between the phase diagrams in Figs.~\ref{PD1d} and \ref{PDGraser} are the same.
Also at $n=5$ the physics is similar to that reported in Fig.~\ref{PDn51d} (not shown).

\subsection{Electronic densities $n=5,6$ with (truncated to 2D) TlFeSe$_2$ hoppings}

\begin{figure}[thbp]
\begin{center}
\includegraphics[width=9.6cm,clip,angle=0]{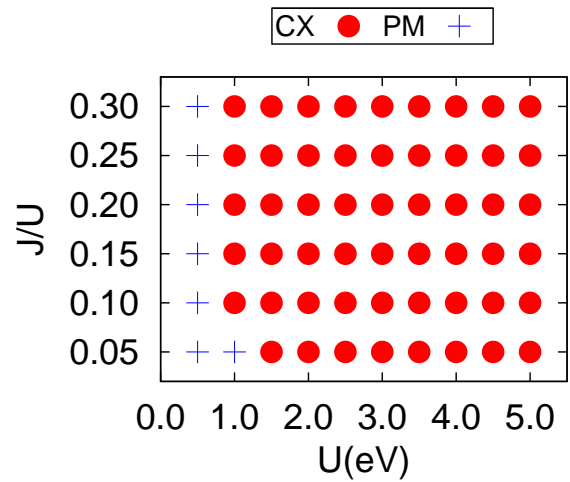}
\caption{Magnetic phase diagram obtained using the real-space 
HF approximation,
an $8 \times 8$ two-dimensional square cluster, and working at a fixed
electronic density $n = 5.0$. 
The hopping selection was explained in the text. The explicit hopping amplitudes 
used for this 2D square lattice includes all the hopping matrices presented 
in the Appendix. The CX state is shown in Fig.~\ref{PD2d}~(b): it has staggered
AFM order along the chains, supplemented by ferromagnetic coupling along
the perpendicular direction with the weaker hoppings. The PM state was obtained under 
the same convention as in Fig.~\ref{PD1d}. The bandwidth at $U=J=0$ is $W = 3.45$~eV.}
\label{PDn52d}
\end{center}
\end{figure}

As explained before, TlFeSe$_2$ and perhaps other quasi one-dimensional materials,
appear to have a temperature range where one dimensionality prevails. However, there
are small couplings in the direction perpendicular to the chain that will eventually
lead to three-dimensional (3D) magnetic order at low temperatures, instead of
merely power-law decaying spin correlations. Studying directly the
full 3D system would be too complicated computationally, due to the rapid growth
with the size of the clusters of the required CPU time 
and the need to have a robust linear 
cluster size to fit the complex spin patterns emerging in our calculations. However, 
an intermediate situation can be achieved if the 1D results are 
merely extended to two dimensions (2D) as opposed to 3D. In practice,
of the vast set of hopping amplitudes obtained from the band structure calculations,
those located within a single two-dimensional layer, 
as those shown in Fig.~\ref{F.cells},
were kept and the rest discarded. Under these circumstances, the real-space 
HF procedure was
repeated, although this time the linear size had to be reduced to keep the computational
time requirements under control. For this reason, an 8$\times$8 cluster was chosen 
for this portion of the study.

For the case of $n = 5.0$, the phase diagram is once again dominated by a single
state as it happened at this electronic density in one dimension. However, the
dominant magnetic order is not a fully staggered G-type AFM state but instead
a CX state with ferromagnetic order along the direction of the weak hoppings. This
is conceptually interesting since it unveils deviations from a simple spin localized
picture due to the presence of many orbitals and the proximity of regions with
double exchange physics.

\begin{figure}[thbp]
\begin{center}
\includegraphics[width=8.0cm,clip,angle=0]{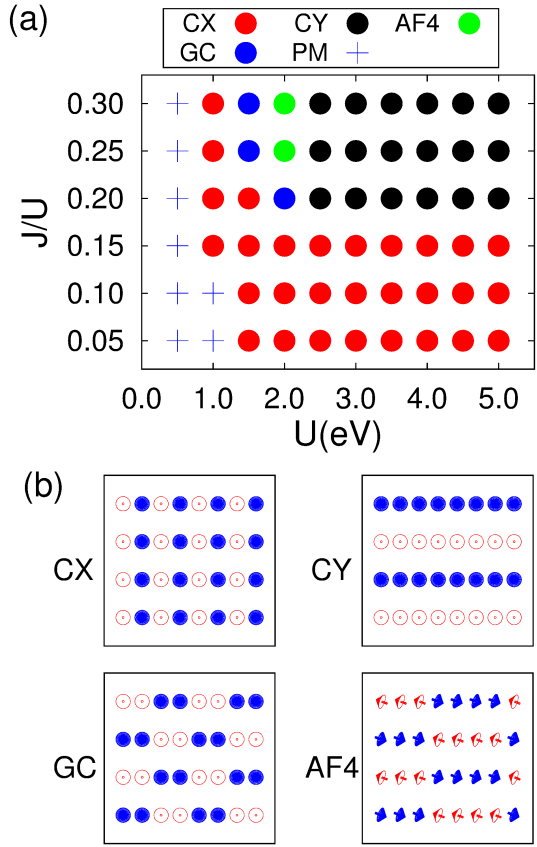}
\caption{(a) Phase diagram obtained using the real-space HF approximation and 
an $8 \times 8$ two-dimensional square lattice, 
at a fixed electronic density $n = 6.0$. 
The hopping selection was explained in the text. The explicit hopping amplitudes 
used for this 2D square lattice includes all the hopping matrices presented 
in the Appendix section. The PM state was obtained under 
the same convention as in Fig.~\ref{PD1d}. The bandwidth at $U=J=0$ is $W = 3.45$~eV.
(b) The definition of the states used in (a). The chains direction is
horizontal.}
\label{PD2d}
\end{center}
\end{figure}

The results of this two-dimensional study for the case of $n=6.0$ 
are shown in Fig.~\ref{PD2d}~(a,b). As in the case of one dimension, the
phase diagram is far richer at $n=6.0$ than at $n=5.0$.
The phase diagram Fig.~\ref{PD2d} 
still displays the small $U$ paramagnetic regime as expected,
followed by the CX state 
for $J/U$ less than 0.20 and any $U$. This state has
staggered AFM order along the 1D dominant direction, but it is FM in the direction
perpendicular to the chains, as already explained for $n=5.0$. 
Thus, this state has C-type AFM characteristics, compatible
with the tendency toward this type of states at $n$=6 in the isotropic layered 
two-dimensional investigations. 
However, at $J/U = 0.20$ or larger, other states appear 
in the phase diagram with increasing $U$, similarly as
in the previously discussed 1D cases. In particular, the FM state of 1D is replaced
by the CY state, where there is ferromagnetism in the dominant chain directions, but
AFM order in the perpendicular direction. This is another 
representative of the C-type AFM
family. It is remarkable that neither the fully staggered G-AFM state nor the fully FM
state are stabilized in the phase diagram in the range of couplings studied, 
but instead in both cases a C-AFM is stable.
Finally, in between the CX and CY states, once again other complex 
states are stabilized,
indicated as GC and AF4 in Fig.~\ref{PD2d}~(b). Here, along the chains direction
these states resemble the Block states AF2 and AF4 of Fig.~\ref{PD1d}, but in the direction perpendicular
the coupling is AFM ({\it Note:} the label GC is used here to match the notation
employed in a recent investigation of two dimensional models varying the electronic density.\cite{luo2d} 
In that context the GC state was found in the phase diagram and the GC label 
denotes a mixture of the G-AFM and C-AFM states). Perhaps
due to limitations in the cluster size studied in this effort, 
other possible states are not present in the phase diagram, but
it is possible that the size of the individual FM clusters 
does not stop at just 4 but a cascade of block states with 
increasing cluster sizes could also
be stabilized (in increasingly narrower regions of parameter space) 
in the transition from AFM to FM states.
Overall, it appears that neither fully FM nor fully staggered AFM states are stabilized,
but instead the phase diagram is dominated by intermediate magnetic states such as the 
C-type AFM and the FM Blocks coupled AFM in both directions.

\section{Density of States and Orbital Composition}

\begin{figure}[thbp]
\begin{center}
\includegraphics[width=8.0cm,clip,angle=0]{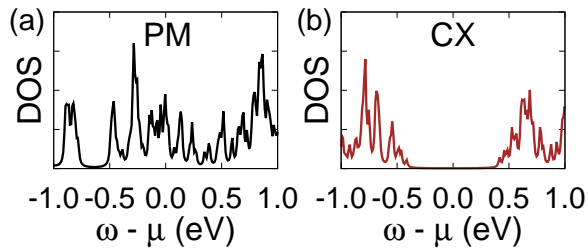}
\caption{ Density of states at representative 
values of couplings, corresponding to the two magnetic states shown in the
2D phase diagram corresponding to $n=5.0$ (Fig.~\ref{PDn51d}). 
(a) PM case $U=J=0.0$; 
(b) CX state at $U=1.0$ and $J/U=0.25$. Increasing further $U$ results
in a concomitant increase in the gap shown in panel (b).}
\label{DOSn5}
\end{center}
\end{figure}

After the real-space HF state is obtained via the iterative process,
then a variety of observables can be calculated. In this section, 
the density of states (DOS) and orbital composition are provided. 
Figure~\ref{DOSn5} contains the DOS for $n=5.0$ in the case of the
2D cluster (see phase diagram in Fig.~\ref{PDn52d}). Although the magnetic
state CX that dominates the phase diagram is not spin staggered in all
directions, the chains are certainly dominating since the hoppings are
the largest along the direction they define. Then, it is not surprising that
the staggered spin order along the chains opens a gap in the DOS 
similarly as it typically occurs in layered G-type AFM states.

\begin{figure}[thbp]
\begin{center}
\includegraphics[width=8.0cm,clip,angle=0]{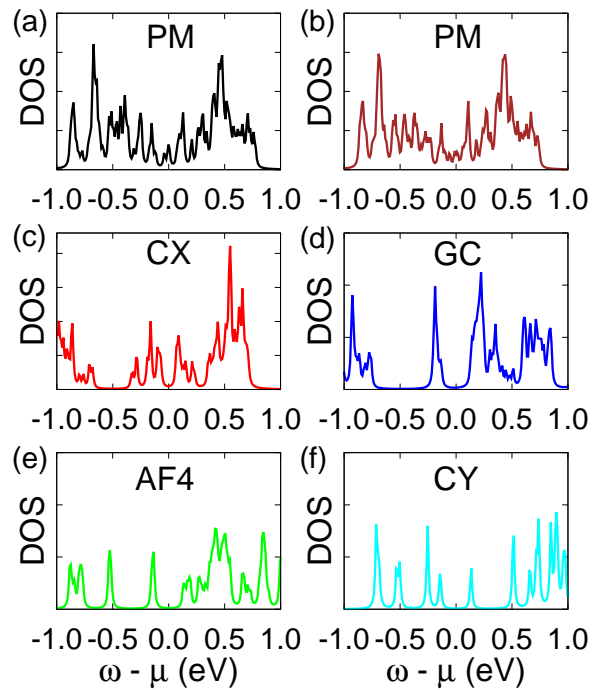}
\caption{ Density of States (DOS) at representative values of couplings, 
corresponding to the five magnetic states shown in 
the 2D phase diagram of $n=6$ (Fig.~\ref{PD2d}). 
(a) PM: $U=J=0.0$; 
(b) PM: $U=0.5$, $J/U=0.25$; (c) CX: $U=1.0$, $J/U=0.25$;
(d) GC: $U=1.5$, $J/U=0.25$; (e) AF4: $U=2.0$, $J/U=0.25$;
(f) CY: $U=3.0$, $J/U=0.25$.}
\label{DOS}
\end{center}
\end{figure}

For the case of $n=6.0$, and the concomitant phase diagram in 2D
shown in Fig.~\ref{PD2d}, the DOS's are displayed in Fig.~\ref{DOS}.
The cases (a) and (b) correspond to the paramagnetic state. Here
the nonzero $U$ and $J$ in (b) induce small modifications in the relative
orbital populations as compared with (a), 
and for this reason these panels are very
similar although not identical. But in both cases a metallic state is observed
as expected (albeit with low weight at the Fermi level, signaling a possible bad metal).
The CX state in panel (c) is close to the border of the paramagnetic state
and for this reason the gap is small of only $\sim$0.1~eV.
Panel (d) contains results for the phase GC.
Since in both directions electrons will have difficulty in propagating
due to the staggered order, and also because $U$ is larger in panel (d)
than in (c), then the gap of the GC state is larger by a factor $\sim$2
than for the CX state. A similar reasoning holds for the AF4 state shown
in panel (e) with a similar gap. Finally, the CY state shown in panel (f) also has a gap
of similar magnitude as the rest, in spite of the fact that panel (f) is
at $U=3$~eV. The reason is that this state has ferromagnetic tendencies
along the chain direction that favor metallicity, and the combination
of this effect together with a robust $U$ conspire to provide a similar
gap as the rest. Summarizing, in a reasonable range of values of $U$ and at
the realistic $J/U=0.25$, the gaps of the many phases 
are approximately similar and in the range 0.1~eV to 0.2~eV.

\begin{figure}[thbp]
\begin{center}
\includegraphics[width=7.0cm,clip,angle=0]{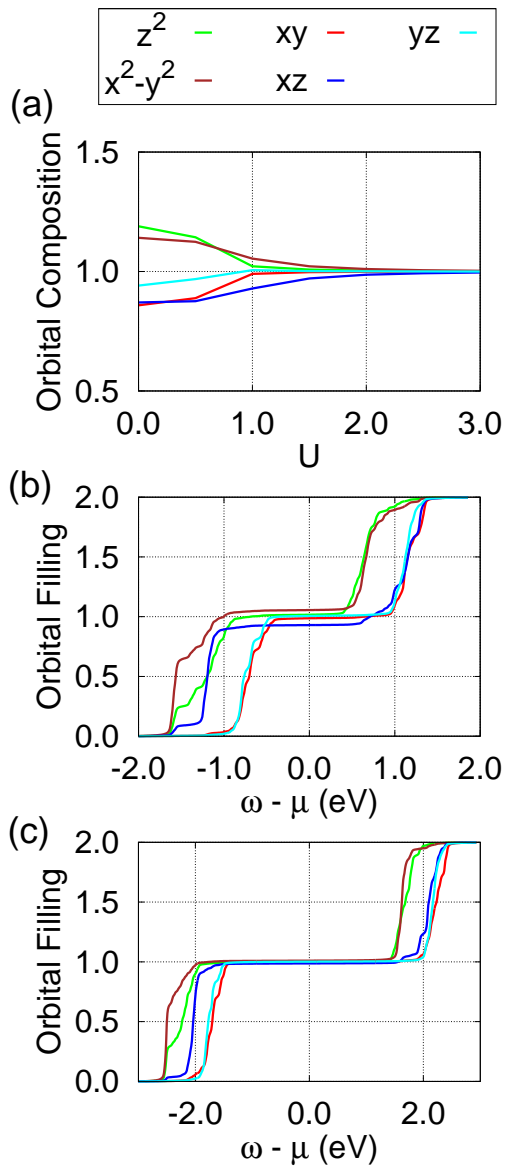}
\caption{ Orbital composition at representative values of the 
couplings in the 2D phase diagram corresponding to electronic density $n=5$. 
(a) Orbital composition vs. $U$ at fixed $J/U=0.25$;
(b) Orbital fillings vs. $\omega - \mu$ at $U=1.0$~eV, and $J/U=0.25$;
(c) Orbital fillings vs. $\omega - \mu$ at $U=2.0$~eV, and $J/U=0.25$.}
\label{Orbn5}
\end{center}
\end{figure}

Figure~\ref{Orbn5} contains information about the orbital composition and filling
for the case of $n=5.0$ and 2D. Panel (a) shows that with increasing $U$, the population
of all the five orbitals converges to 1, even though at $U=0.0$ there is an energy
splitting. This is in agreement with expectations at this electronic density.
At $U=1.0$~eV, panel (a) shows that still the individual orbital population is not
precisely 1. Then, panel (b) 
does not have a sharp gap near 0.0. But with increasing Hubbard coupling, panel (c)
displays a robust gap-like feature at $U=2.0$~eV for all the five curves 
since at this  coupling all the orbitals are already in the Mott regime.

\begin{figure}[thbp]
\begin{center}
\includegraphics[width=7.0cm,clip,angle=0]{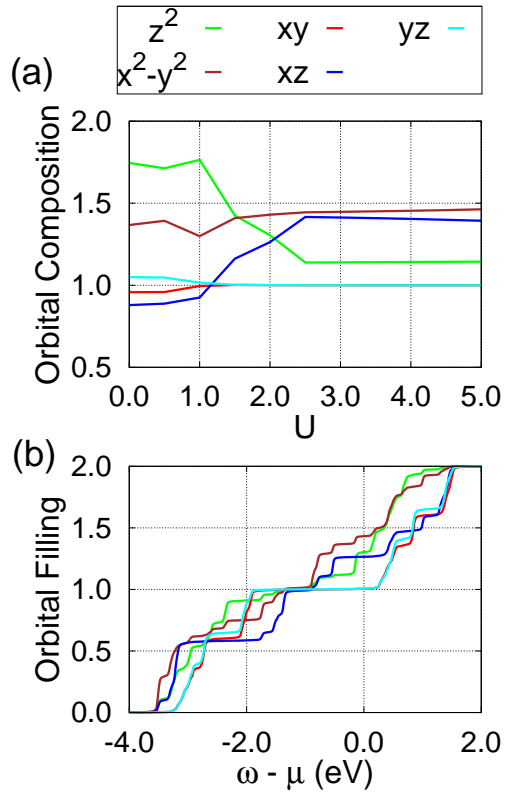}
\caption{ Orbital composition at representative values 
of couplings in the 2D phase diagram. (a) Orbital composition vs. $U$, at fixed $J/U=0.25$;
(b) Orbital fillings vs. $\omega - \mu$ at $U=2.0$, $J/U=0.25$. 
The total electron filling is $n=6.0$ for both (a) and (b).}
\label{Orb}
\end{center}
\end{figure}

The case of $n=6.0$ is quite different from $n=5.0$. Figure~\ref{Orb}~(a)
contains the orbital composition with increasing $U$, at $J/U=0.25$. Compatible
with the ideas related with orbital selective Mott transitions,\cite{Georges} in this
case there are two orbitals that converge to population 1 in the range of $U$ between 1 and 2, 
while the rest still carry a non-integer number of electrons. 
As a consequence, this is a state that combines
localized features related with the orbitals in the Mott state and itinerant
features related with the rest of the orbitals. Previous mean-field calculations
in layered 2D systems (isotropic in space) have also reported the presence of
a similar orbital-selective state.\cite{previousOSMP} 
This is also compatible with the relatively
small gaps observed in the DOS at $n=6.0$ as compared with $n=5.0$.
Figure~\ref{Orb}~(b) illustrates these points once again: the two orbitals with
population 1 have a robust gap signaled by the plateaux in the curve, while
the rest only have a smaller gap caused by the influence of the localized spins
on the itinerant carriers. Thus, the cases of $n=5.0$ and $6.0$ are fundamentally
different.

\section{Conclusions}

In this publication, an electronic model Hamiltonian for 
one-dimensional iron-selenides was studied. The phase
diagrams were constructed at the $3d$ iron electronic densities
$n=5.0$ and $6.0$, using the numerically demanding real-space 
Hartree Fock approximation. The results at $n=5.0$ were obtained
employing hopping amplitudes that were derived using $ab-initio$ techniques 
from the already synthesized material TlFeSe$_2$ with quasi one-dimensional structures.
The $n=5.0$ phase diagram is dominated by staggered spin
order along the dominant chains direction, 
and a robust gap in the density of states. 

The case of $n=6.0$ does not correspond to any known one-dimensional selenide compound
to our knowledge, thus our results have the goal of motivating experimental groups for the
preparation of such a material. Our study unveils a phase
diagram far richer than at $n=5.0$, particularly at the often quoted ratio $J/U=0.25$
for the iron superconductors. In this regime, the $n=6.0$ phase diagram contains
a variety of novel states involving block phases with clusters of ferromagnetic
spins of various lengths, antiferromagnetically coupled among them, as in
recent DMRG studies by Rinc\'on {\it et al.}\cite{rincon} The density of
states reveals a relatively small gap that may be indicative of a weak insulator or 
bad metallic behavior in a real material. Note, however, that our study was carried out
employing the hopping amplitudes of the case $n=5.0$. Then, if a $n=6.0$ material is
ever synthesized then the present calculations should be redone with more realistic 
hoppings derived from 
density functional theory. Nevertheless, results at $n=5.0$ also shown in our
present effort indicate a weak dependence on the actual hopping set as long as 
a particular direction dominates, suggesting that the block states may exist
in real $n=6.0$ quasi one-dimensional compounds. Considering the importance of
similar unidimensional materials in the context of the cuprates together with our present analysis,
all indicates that carrying out low dimensional investigations in the framework
of iron selenides and pnictides may provide valuable information about the
widely studied iron-based superconductors.

\section{Acknowledgment}

We thank Juli\'an Rinc\'on for useful conversations.
The work of the authors was supported by the U.S. Department of Energy,
Office of Basic Energy Sciences, Materials Sciences and
Engineering Division. 

\section{Appendix}

\begin{figure}[tb]
\begin{center}
\includegraphics[trim = 0mm 0mm 0mm 0mm, clip,width=1.0\columnwidth]{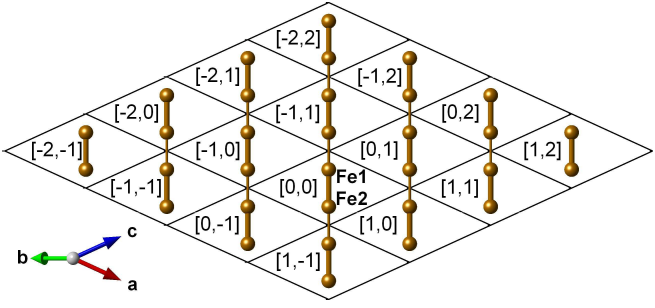}
\end{center}
\caption{Schematic representation of the {\tlfese} crystal structure
with only the Fe atoms in the $ac$ plane shown. The Fe chains run along
the $-\vec{a}+\vec{c}$ direction. The shorter, 2.74~\AA, (longer, 2.75~\AA)
NN distances between the Fe atoms are indicated by thick (thin) sticks connecting the atoms.
$[x,z]$ labels the unit cells.
}
\label{F.cells}
\end{figure}

For completeness, a list of the electron hopping
amplitudes between the Fe~$3d$ orbitals in the
$\{d_{z^2},d_{x^2-y^2},d_{xy},d_{xz},d_{yz}\}$
orbital basis is here provided. The matrix $t^{1j}_{[x,z]}$ describes the hopping amplitudes
between $3d$ orbitals of Fe atom $1$ in the unit cell
$[0,0]$ and those of Fe atom $j$ in the unit cell $[x,z]$,
as shown in Fig.~\ref{F.cells}.

\begin{center}
{\small
\begin{tabular}{p{1.5cm}p{6.9cm}}
\hline \hline\noalign{\smallskip}
Matrix & \hspace{2.9cm}{\tlfese}\\ \hline\noalign{\smallskip}
$t^{11}_{[0,0]}$ &
$\begin{pmatrix}
 \textendash0.2643 & \textendash0.0859 &  0      & \textendash0.0070 &  0     \\
 \textendash0.0859 & \textendash0.3711 &  0      & ~~0.0127&  0     \\
  0      &  0      & ~~0.2926&  0      & \textendash0.0089\\
 \textendash0.0070 & ~~0.0127&  0      & ~~0.0882&  0     \\
  0      &  0      & \textendash0.0089 &  0      & ~~0.2427
\end{pmatrix}$
\end{tabular}
\begin{tabular}{p{1.5cm}p{6.9cm}}
\hline
$\begin{array}{l}
t^{12}_{[0,0]}\\\\
d\text{=2.74~\AA}
\end{array}$
&
$\begin{pmatrix}
 ~~0.0405& \textendash0.2980 &  0      & \textendash0.2630 &  0     \\
 \textendash0.2980 & \textendash0.2376 &  0      & \textendash0.4202 &  0     \\
  0      &  0      & \textendash0.2253 &  0      & \textendash0.2593\\
 \textendash0.2630 & \textendash0.4202 &  0      & \textendash0.4332 &  0     \\
  0      &  0      & \textendash0.2593 &  0      & \textendash0.2550
\end{pmatrix}$
\end{tabular}
\begin{tabular}{p{1.5cm}p{6.9cm}}
\hline
$\begin{array}{l}
t^{12}_{[-1,1]}\\\\
d\text{=2.75~\AA}
\end{array}$
&
$\begin{pmatrix}
 ~~0.0901& \textendash0.2739 &  0      & ~~0.2164&  0     \\
 \textendash0.2739 & \textendash0.2459 &  0      & ~~0.4549&  0     \\
  0      &  0      & \textendash0.2884 &  0      & ~~0.2356\\
 ~~0.2164& ~~0.4549&  0      & \textendash0.4450 &  0     \\
  0      &  0      & ~~0.2356&  0      & \textendash0.2156
\end{pmatrix}$
\end{tabular}
\begin{tabular}{p{1.5cm}p{6.9cm}}
\hline
$\begin{array}{l}
t^{11}_{[-1,1]}\\\\
d\text{=5.49~\AA}
\end{array}$
&
$\begin{pmatrix}
 \textendash0.0695 & ~~0.0184&  0      & \textendash0.0128 &  0     \\
 ~~0.0062& \textendash0.0037 &  0      & \textendash0.0322 &  0     \\
  0      &  0      & \textendash0.0426 &  0      & \textendash0.0130\\
 ~~0.0139& ~~0.0399&  0      & \textendash0.0416 &  0     \\
  0      &  0      & ~~0.0157&  0      & \textendash0.0202
\end{pmatrix}$
\end{tabular}
\begin{tabular}{p{1.5cm}p{6.9cm}}
\hline
$\begin{array}{l}
t^{11}_{[1,-1]}\\\\
d\text{=5.49~\AA}
\end{array}$
&
$\begin{pmatrix}
 \textendash0.0695 & ~~0.0062&  0      & ~~0.0139&  0     \\
 ~~0.0184& \textendash0.0037 &  0      & ~~0.0399&  0     \\
  0      &  0      & \textendash0.0426 &  0      & ~~0.0157\\
 \textendash0.0128 & \textendash0.0322 &  0      & \textendash0.0416 &  0     \\
  0      &  0      & \textendash0.0130 &  0      & \textendash0.0202
\end{pmatrix}$
\end{tabular}
\begin{tabular}{p{1.5cm}p{6.9cm}}
\hline
$\begin{array}{l}
t^{12}_{[-1,0]}\\\\
d\text{=5.99~\AA}
\end{array}$
&
$\begin{pmatrix}
\textendash0.0601& \textendash0.0075& \textendash0.0004& \textendash0.0020& \textendash0.0001\\
\textendash0.0075& ~~0.0085& ~~0.0021& \textendash0.0047& \textendash0.0115\\
\textendash0.0004& ~~0.0021& ~~0.0152& ~~0.0099& \textendash0.0210\\
\textendash0.0020& \textendash0.0047& ~~0.0099& \textendash0.0029& ~~0.0164\\
\textendash0.0001& \textendash0.0115& \textendash0.0210& ~~0.0164& \textendash0.0521
\end{pmatrix}$
\end{tabular}
\begin{tabular}{p{1.5cm}p{6.9cm}}
\hline
$\begin{array}{l}
t^{12}_{[0,1]}\\\\
d\text{=5.99~\AA}
\end{array}$
&
$\begin{pmatrix}
\textendash0.0601& \textendash0.0075& ~~0.0004& \textendash0.0020& ~~0.0001\\
\textendash0.0075& ~~0.0085& \textendash0.0021& \textendash0.0047& ~~0.0115\\
~~0.0004& \textendash0.0021& ~~0.0152& \textendash0.0099& \textendash0.0210\\
\textendash0.0020& \textendash0.0047& \textendash0.0099& \textendash0.0029& \textendash0.0164\\
~~0.0001& ~~0.0115& \textendash0.0210& \textendash0.0164& \textendash0.0521
\end{pmatrix}$
\end{tabular}
\begin{tabular}{p{1.5cm}p{6.9cm}}
\hline
$\begin{array}{l}
t^{11}_{[0,-1]}\\\\
d\text{=6.59~\AA}
\end{array}$
&
$\begin{pmatrix}
\textendash0.0286& ~~0.0362& \textendash0.0083& \textendash0.0055& \textendash0.0197\\
\textendash0.0075& ~~0.0147& ~~0.0064& ~~0.0199& ~~0.0265\\
\textendash0.0527& \textendash0.0049& \textendash0.0048& ~~0.0167& \textendash0.0193\\
~~0.0220& ~~0.0217& \textendash0.0156& ~~0.0079& ~~0.0100\\
\textendash0.0241& ~~0.0067& ~~0.0094& ~~0.0195& ~~0.0127
\end{pmatrix}$
\end{tabular}
\begin{tabular}{p{1.5cm}p{6.9cm}}
\hline
$\begin{array}{l}
t^{11}_{[1,0]}\\\\
d\text{=6.59~\AA}
\end{array}$
&
$\begin{pmatrix}
\textendash0.0286& ~~0.0362& ~~0.0083& \textendash0.0055& ~~0.0197\\
\textendash0.0075& ~~0.0147& \textendash0.0064& ~~0.0199& \textendash0.0265\\
~~0.0527& ~~0.0049& \textendash0.0048& \textendash0.0167& \textendash0.0193\\
~~0.0220& ~~0.0217& ~~0.0156& ~~0.0079& \textendash0.0100\\
~~0.0241& \textendash0.0067& ~~0.0094& \textendash0.0195& ~~0.0127
\end{pmatrix}$
\end{tabular}
\begin{tabular}{p{1.5cm}p{6.9cm}}
\hline
$\begin{array}{l}
t^{11}_{[-1,0]}\\\\
d\text{=6.59~\AA}
\end{array}$
&
$\begin{pmatrix}
\textendash0.0286& \textendash0.0075& ~~0.0527& ~~0.0220& ~~0.0241\\
~~0.0362& ~~0.0147& ~~0.0049& ~~0.0217& \textendash0.0067\\
~~0.0083& \textendash0.0064& \textendash0.0048& ~~0.0156& ~~0.0094\\
\textendash0.0055& ~~0.0199& \textendash0.0167& ~~0.0079& \textendash0.0195\\
~~0.0197& \textendash0.0265& \textendash0.0193& \textendash0.0100& ~~0.0127
\end{pmatrix}$
\end{tabular}
\begin{tabular}{p{1.5cm}p{6.9cm}}
\hline
$\begin{array}{l}
t^{11}_{[0,1]}\\\\
d\text{=6.59~\AA}
\end{array}$
&
$\begin{pmatrix}
\textendash0.0286& \textendash0.0075& \textendash0.0527& ~~0.0220& \textendash0.0241\\
~~0.0362& ~~0.0147& \textendash0.0049& ~~0.0217& ~~0.0067\\
\textendash0.0083& ~~0.0064& \textendash0.0048& \textendash0.0156& ~~0.0094\\
\textendash0.0055& ~~0.0199& ~~0.0167& ~~0.0079& ~~0.0195\\
\textendash0.0197& ~~0.0265& \textendash0.0193& ~~0.0100& ~~0.0127
\end{pmatrix}$
\end{tabular}
\begin{tabular}{p{1.5cm}p{6.9cm}}
\hline
$\begin{array}{l}
t^{12}_{[1,0]}\\\\
d\text{=8.12~\AA}
\end{array}$
&
$\begin{pmatrix}
~~0.0097& ~~0.0150& ~~0.0066& ~~0.0158& \textendash0.0061\\
~~0.0150& \textendash0.0150& \textendash0.0267& \textendash0.0127& \textendash0.0089\\
~~0.0066& \textendash0.0267& \textendash0.0609& \textendash0.0573& \textendash0.0130\\
~~0.0158& \textendash0.0127& \textendash0.0573& \textendash0.0272& \textendash0.0058\\
\textendash0.0061& \textendash0.0089& \textendash0.0130& \textendash0.0058& ~~0.0181
\end{pmatrix}$
\end{tabular}
\begin{tabular}{p{1.5cm}p{6.9cm}}
\hline
$\begin{array}{l}
t^{12}_{[0,-1]}\\\\
d\text{=8.12~\AA}
\end{array}$
&
$\begin{pmatrix}
~~0.0097& ~~0.0150& \textendash0.0066& ~~0.0158& ~~0.0061\\
~~0.0150& \textendash0.0150& ~~0.0267& \textendash0.0127& ~~0.0089\\
\textendash0.0066& ~~0.0267& \textendash0.0609& ~~0.0573& \textendash0.0130\\
~~0.0158& \textendash0.0127& ~~0.0573& \textendash0.0272& ~~0.0058\\
~~0.0061& ~~0.0089& \textendash0.0130& ~~0.0058& ~~0.0181
\end{pmatrix}$
\end{tabular}
\begin{tabular}{p{1.5cm}p{6.9cm}}
\hline
$\begin{array}{l}
t^{12}_{[-2,1]}\\\\
d\text{=8.13~\AA}
\end{array}$
&
$\begin{pmatrix}
~~0.0049& ~~0.0123& \textendash0.0125& \textendash0.0034& \textendash0.0123\\
~~0.0123& \textendash0.0094& ~~0.0091& ~~0.0130& \textendash0.0033\\
\textendash0.0125& ~~0.0091& ~~0.0074& ~~0.0114& ~~0.0142\\
\textendash0.0034& ~~0.0130& ~~0.0114& ~~0.0067& ~~0.0078\\
\textendash0.0123& \textendash0.0033& ~~0.0142& ~~0.0078& \textendash0.0002
\end{pmatrix}$
\end{tabular}
\begin{tabular}{p{1.5cm}p{6.9cm}}
\hline
$\begin{array}{l}
t^{12}_{[-1,2]}\\\\
d\text{=8.13~\AA}
\end{array}$
&
$\begin{pmatrix}
~~0.0049& ~~0.0123& ~~0.0125& \textendash0.0034& ~~0.0123\\
~~0.0123& \textendash0.0094& \textendash0.0091& ~~0.0130& ~~0.0033\\
~~0.0125& \textendash0.0091& ~~0.0074& \textendash0.0114& ~~0.0142\\
\textendash0.0034& ~~0.0130& \textendash0.0114& ~~0.0067& \textendash0.0078\\
~~0.0123& ~~0.0033& ~~0.0142& \textendash0.0078& \textendash0.0002
\end{pmatrix}$
\end{tabular}
\begin{tabular}{p{1.5cm}p{6.9cm}}
\hline
$\begin{array}{l}
t^{12}_{[1,-1]}\\\\
d\text{=8.23~\AA}
\end{array}$
&
$\begin{pmatrix}
~~0.0306& \textendash0.0140&  0     & ~~0.0156&  0     \\
\textendash0.0140& ~~0.0053&  0     & \textendash0.0054&  0     \\
 0     &  0     & ~~0.0042&  0     & \textendash0.0126\\
~~0.0156& \textendash0.0054&  0     & ~~0.0140&  0     \\
 0     &  0     & \textendash0.0126&  0     & \textendash0.0456
\end{pmatrix}$
\end{tabular}
\begin{tabular}{p{1.5cm}p{6.9cm}}
\hline
$\begin{array}{l}
t^{12}_{[-2,2]}\\\\
d\text{=8.24~\AA}
\end{array}$
&
$\begin{pmatrix}
~~0.0272& \textendash0.0082&  0     & ~~0.0025&  0     \\
\textendash0.0082& \textendash0.0037&  0     & ~~0.0169&  0     \\
 0     &  0     & \textendash0.0082&  0     & ~~0.0240\\
~~0.0025& ~~0.0169&  0     & ~~0.0052&  0     \\
 0     &  0     & ~~0.0240&  0     & \textendash0.0373
\end{pmatrix}$\\
\hline\hline
\end{tabular}
}
\end{center}

\vspace*{5.55cm}


\end{document}